\documentclass[preprint,noshowpacs,preprintnumbers,amsmath,amssymb,superscriptaddress]{revtex4-1}
\usepackage{graphicx}
\begin{document}
\title{Gated Silicene as a tunable source of nearly 100\% spin-polarized electrons}

\author{Wei-Feng Tsai}
\affiliation{Department of Physics, National Sun Yat-sen University, Kaohsiung 80424, Taiwan}
\author{Cheng-Yi Huang}
\affiliation{Department of Physics, National Sun Yat-sen University, Kaohsiung 80424, Taiwan}
\author{Tay-Rong Chang}
\affiliation{Department of Physics, National Tsing Hua University, Hsinchu 30013, Taiwan}
\author{Hsin Lin$^*$}
\affiliation{Department of Physics, Northeastern University, Boston, Massachusetts 02115, USA}
\author{Horng-Tay Jeng}
\affiliation{Department of Physics, National Tsing Hua University, Hsinchu 30013, Taiwan}
\affiliation{Institute of Physics, Academia Sinica, Taipei 11529, Taiwan}
\author{A. Bansil}
\affiliation{Department of Physics, Northeastern University, Boston, Massachusetts 02115, USA}

\date{\today}

\begin{abstract}
{\bf Silicene is a one-atom-thick 2D crystal of silicon with a hexagonal lattice structure that is related to that of graphene but with atomic bonds that are buckled rather than flat. This buckling confers advantages on silicene over graphene, because it should, in principle, generate both a band gap and polarized spin-states that can be controlled with a perpendicular electric field. Here we use first-principles calculations to show that field-gated silicene possesses two gapped Dirac cones exhibiting nearly 100\% spin-polarization, situated at the corners of the Brillouin zone. Using this fact, we propose a design for a silicene-based spin-filter that should enable the spin-polarization of an output current to be switched electrically, without switching external magnetic fields. Our quantum transport calculations indicate that the proposed designs will be highly efficient (nearly 100\% spin polarization) and robust against weak disorder and edge imperfections. We also propose a Y-shaped spin/valley separator that produces spin-polarized current at two output terminals with opposite spins.

}

\end{abstract}
\maketitle

\section{Introduction}

Controllable non-magnetic spintronic devices are highly desirable in the development of spintronics applications which is an area of intense current interest\cite{Awschalom1,Awschalom2,spintronics}.
While the spin polarization is generated by spin-orbit coupling (SOC),
the entanglement between spin and orbital degrees of freedom due to the SOC significantly reduces the degree of spin polarization of spin-split states in most non-magnetic semiconductors including the topological insulators \cite{LouieTIspin}.
Recently, silicene, a close relative of graphene \cite{review-graphene},
 has been predicted and synthesized \cite{honeycomb1,honeycomb2,honeycomb3,silicene1,silicene2,silicene3,silicene4,
silicene5,silicene6,silicene7,silicene8}.
 The band structure of silicene is similar to that of graphene in that the conduction and valence edges occur at the corners (K and K' points) of the Brillouin zone. Silicene however possesses fundamental differences from graphene driven by the presence of a larger SOC, which opens gaps at the K-points \cite{liu11a,liu11b}. These gaps can be tuned with an external E-field perpendicular to the plane, which breaks the inversion symmetry (IS) of the system due to the presence of buckling in the honeycomb structure \cite{drummond12}. In this way, silicene can overcome difficulties associated with graphene in electronics applications (lack of a controllable gap) \cite{graphenegap1,graphenegap2}, potential applications of graphene in nano-electronics due to the available spin, valley and pseudo-spin degrees of freedom notwithstanding \cite{valley1,valley2,valley3,valley4,mos2,mos2prb}.

In this work we present detailed first-principles computations to show that the band structure of gated silicene harbors two nearly 100\% spin-polarized Dirac cones at the K-points. We use this key finding to propose designs of a silicene-based spin-filter as well as a spin-separator.
Our quantum transport calculations show how a silicene based high efficiency spin filter (nearly 100\% spin polarization) suitable for high-frequency electronics applications can be designed to switch the output spin current simply by gating without the need to switch magnetic domains. We analyze reflection and transmission of current at the junction of two domains with different external fields in order to obtain insight into transport selectivity between spin and valley degrees of freedom. High efficiency and tunability of our spin filter takes advantage of bulk charge carriers rather than the edge current in quantum spin Hall (QSH) systems, making our silicene filter robust against weak disorder and edge imperfections.
We also propose a Y-shaped spin/valley separator that produces spin-polarized current at two output terminals with opposite spins. This three terminal device could be used for logical circuits beyond binary operations. Ge, Sn, and Pb counterparts of silicene are shown to have similar properties, but their larger SOC results in larger energy differences between the spin-split states making these materials better suited for room temperature applications.
Silicene, germanene, Sn and Pb thin films would thus be ideal materials for functional electronics and spintronics applications. Silicene however stands out because it could be integrated more naturally into the existing Si-based technologies.


\section{Results}

{\bf Band structure and phase transition via IS breaking.}
Fig.~\ref{Fig1} shows the low-buckled honeycomb lattice of silicene with its two sublattices A and B, and key features of the band structure. At low energies, the $\pi$-electrons dictate the system and reside around two inequivalent valleys, K and K' points, in the first Brillouin zone. Although we have carried out first-principles computations, insight into the underlying physics is obtained by describing the system by a minimal, four-band tight-binding model \cite{liu11a,liu11b}
\begin{eqnarray}
H &=& -t\sum_{\langle ij\rangle,\sigma}c^\dagger_{i\sigma}c_{j\sigma}
+i\frac{\lambda_{\text{SO}}}{3\sqrt{3}}\sum_{\langle\langle ij\rangle\rangle,\sigma\sigma^\prime}\nu_{ij}c^\dagger_{i\sigma}
\bold{s}^{z}_{\sigma\sigma^\prime}c_{j\sigma^\prime} \nonumber \\
&+& i\frac{2\lambda_{\text{R}}}{3}\sum_{\langle\langle ij\rangle\rangle,\sigma\sigma^\prime}
\mu_{ij}c^\dagger_{i\sigma}
(\bold{s}\times\bold{\hat{d}}_{ij})^{z}_{\sigma\sigma^\prime}c_{j\sigma^\prime}
\nonumber \\
&+& \lambda_{\text{v}}\sum_{i,\sigma}\xi_{i}c^\dagger_{i\sigma}c_{i\sigma}+
h\sum_{i,\sigma\sigma^\prime}c^\dagger_{i\sigma}\bold{s}^z_{\sigma\sigma^\prime}
c_{i\sigma^\prime}, \label{eq:hamiltonian}
\end{eqnarray}
where $c^\dagger_{i\sigma}$ creates an electron at site $i$ with spin polarization $\sigma$. The first term is a nearest-neighbor (NN) hopping term on the honeycomb lattice. The second and third terms are intrinsic (IS preserved) and Rashba (IS preserved but $z\rightarrow -z$ mirror symmetry violated) spin-orbit interactions, respectively, which involve spin dependent next nearest-neighbor (NNN) hopping. $\bold{s}$ are the Pauli spin-matrices. $\nu_{ij}=(\bold{d}_i\times\bold{d}_j)^z/|\bold{d}_i\times\bold{d}_j|=\pm 1$ with two NN bonds, $\bold{d}_i$ and $\bold{d}_j$, connecting the NNN $\bold{d}_{ij}$. $\mu_{ij}=1 (-1)$ when connecting A-A (B-B) sites. Notably, these SOCs, originating from buckling of the structure, are what distinguish silicene from graphene, even though both materials form a honeycomb lattice. The fourth term is an IS broken, staggered sublattice potential ($\xi_{i}=\pm 1$), which arises when an external out-of-plane electric field, $E_z$, is applied. The fifth term represents the effect of an applied out-of plane Zeeman (exchange) field with strength $h$. The last two terms, corresponding to the microscopic responses to the external fields, turn out to be the driving forces for the phase transition discussed below. A phase diagram is provided as Supplementary Fig.~S1 in the Supplementary Information. All the coupling parameters (Table I) have been obtained by fitting to first-principles band structures of stand-alone silicene sheet and Ge, Sn, Pb counterparts. 
Note that the fourth term in our Hamiltonian of Eq.~(\ref{eq:hamiltonian}) refers to the net electrostatic potential difference due to the environment, and accounts for effects of both the external field and the presence of the substrate. We expect therefore our modeling to be relevant for silicene placed on a substrate so long as the coupling with the substrate is not so strong as to modify the electronic structure of the silicene sheet. Although the band gap could be enhanced through strong interaction with the substrate, which is desirable for room temperature applications, it will be a tradeoff between achieving a larger gap and the possible reduction in the degree of spin-polarization resulting from the involvement of substrate orbitals. Moreover, the origin of the band gap itself could be different for different substrates. In order to help guide search for a suitable substrate, Supplementary Fig.~S2 gives the size of the band gap as a function of the lattice constant. It would perhaps be best to proceed by placing the silicene sheet on an insulator or semiconductor substrate where the lattice mismatch is not too large. It will be interesting however to further examine strong substrate coupling effects on a silicene sheet.  Finally, we note that if a non-magnetic metal with small spin-orbit coupling is used as contact material, the spin-up as well as spin-down states of silicene will couple with more or less equal strength with the substrate and as a result spin-transport will not be affected much even though contact resistance may be altered substantially.

In the absence of external fields, two of the four bands are occupied in Eq.~(\ref{eq:hamiltonian}), so that the ground state of silicene is a QSH insulator with a SOC gap of $2\lambda_{\text{SO}}$=8.4meV \cite{liu11a,drummond12}. The topological (Z$_2$) nature of this ground state can be easily examined by either the parity analysis\cite{fu07} or by showing the presence of helical edge states in a zigzag strip geometry. Note that the spin degrees of freedom in the band structure are still degenerate, as a consequence of both time reversal symmetry (TRS) and IS, even though $\bold{s}_z$ is no longer a good quantum number in the presence of NNN Rashba SOC (except at K-points where $\lambda_{\text{R}}$ term makes {\it no} contribution).

When $E_z$ is applied to the buckled structure in which the two atoms within the unit cell are not coplanar, we naturally obtain a non-vanishing $\lambda_{\text{v}}=l E_z$ caused by IS breaking. The estimated value of $l$ is given in Table I. At K point, the energy gap, now $2|\lambda_{\text{SO}}-\lambda_{\text{v}}|$, decreases linearly as $E_z$ increases until a critical field $E_{\text{c}}$ is reached. More importantly, the spin degeneracy is lifted with energy splitting of $2\lambda_{\text{v}}$ as shown schematically in Fig.~\ref{Fig1}b and precisely in Fig.~1e. We emphasize that this splitting arises from the spatial IS breaking as well as the presence of SOC in silicene. The spin polarization profile around the K' point is opposite to that of the K point as required by TRS. Furthermore, by symmetry arguments mentioned above, $\bold{s}_z$ remains a good quantum number at K-points.

As approaching the critical field, $E_z=E_{\text{c}}$, the gap shrinks to zero, forming a Dirac-like cone near each valley with the spectrum, $\sqrt{k^2+m^2}$, where $m=\sqrt{\lambda_{\text{v}}^2+a^2\lambda_{\text{R}}^2 k^2}$ and $k$ denotes the momentum with respect to the K-point (See Fig.~\ref{Fig1}c). In sharp contrast to the Dirac cone in graphene, where each cone is spin degenerate, here the spin is fully polarized along the $z$ direction at each valley (Fig.~1d) and develops a small in-plane component of the form, $\alpha(\bold{s}\times\bold{k})^z$,  with $\alpha$ of order $\lambda_{\text{R}}$. Therefore, in the critical phase (dubbed {\it spin-valley-polarization metal} (SVPM)\cite{ezawa12}) near half-filling, the physics of the system is dominated by two nearly fully spin-polarized (with opposite polarizations) Dirac-like cones at K and K' points.

For $E_z>E_{\text{c}}$, the energy gap re-opens at each valley and drives the QSH phase into the topologically trivial band insulating (BI) phase. It is interesting to note that the spin polarization is unchanged around K-point, although the probability amplitudes for A and B sublattices in the wavefunction is different.
Starting from the BI phase with spin polarization of each band similar to that shown in Fig.~\ref{Fig1}b, a Zeeman field $h$ along -$z$ direction cannot affect the {\it direct} band gap at each valley. However, the conduction band bottom would shift downward at K' point, while the valence band top would shift upward at K point. Increasing $h$ therefore reduces the indirect band gap of the system and drives it into a valley-imbalanced metallic state, dubbed {\it valley-polarized metal} (VPM) \cite{ezawa12}, after crossing the zero {\it indirect} gap point.


{\bf Field-tunable spin filter.} Because various phases in silicene can be realized through the interplay of applied electric and magnetic fields \cite{ezawa12}, it becomes possible to set up a silicene based device working as a high efficiency spin filter. We illustrate this possibility by considering a 2D device consisting of a quantum point contact (QPC) in a silicene thin film. As seen clearly in Fig.~\ref{Fig2}a, a QPC is characterized by a short and narrow constriction. When a current $I$ passes through the device by the application of an applied potential across the two wide regions, the conductance is sharply quantized in units of $e^2/h$. To make the valley degrees of freedom well separated and thus keep the features unique, we choose the zigzag edges for the whole geometry along the direction of current flow. Note that a similar device has been proposed to realize a valley filter in graphene, but that works by using special properties of the edge state and it is completely different from our work here \cite{valley2}.

We now show through quantum transport calculations that on the first few conductance plateaus the QPC produces an almost fully spin-polarized current. By locally changing the potential barrier via gating control in the constriction, the spin polarization direction can be easily reversed. To begin with, we map the tight-binding description of silicene, Eq.~(\ref{eq:hamiltonian}), to the geometry shown in Fig.~\ref{Fig2}a. The two opposite wide regions in the SVPM phase are arranged with Fermi energy $E_{\text{F}}=0.07t$ to model metallic source and drain ($E_{\text{F}}=0$ defines position of the ``Dirac point''). Note that in the source region both spin polarizations are present (See Fig.~\ref{Fig2}c). In the constriction, we keep the system in the marginal VPM phase with a fixed $E_z>E_{\text{c}}$ and the applied Zeeman field $h=\lambda_{\text{v}}-\lambda_{\text{SO}}$ such that the indirect gap becomes zero (See Fig.~\ref{Fig2}d). Furthermore, we add an electrostatic potential barrier $U(x_i)$, which varies only along the current flow direction and is non-vanishing only in the constriction region,
\begin{equation}
U(x_i)=U_0[\Theta_{L_{\text{s}}}(x_i)-\Theta_{L_{\text{s}}}(x_i-L)],
\end{equation}
where $L$ is the length of the narrow region, $L_{\text{s}}$ is a parameter for smoothing the potential, $\Theta_{L_{\text{s}}}(x_i)=0$ (1) for $x_i<-L_{\text{s}}/2$ ($x_i>L_{\text{s}}/2$) and $\Theta_{L_{\text{s}}}(x_i)=\frac{1}{2}+\frac{1}{2}\sin(\frac{\pi x_i}{L_{\text{s}}})$ for $|x_i|<L_{\text{s}}/2$ ($x_i=\frac{L}{2}$ defines the middle of the constriction).
Two types of potential barriers are considered: 1) $L_{\text{s}}=0$, a rectangular shape with potential height $U_0$ and 2) $0<L_{\text{s}}<L$, a smooth shape, as shown in Fig. 2a (lower part) by solid and dashed curves respectively.

A typical dispersion relation for the wide region in the SVPM phase is explicitly shown in Fig.~\ref{Fig2}c. Given $E_{\text{F}}$, this region provides a total of $2N$ right-moving modes. In particular, as long as $E_{\text{F}}< (\lambda_{\text{v}}+\lambda_{\text{SO}})$ and $\lambda_{\text{R}}\ll\{\lambda_{\text{SO}},t\}$, $N$ propagating modes coming from the left valley will carry only down-spins and the other half modes coming from the right valley will carry only up-spins. This result has the merit of conveniently counting the spin polarization defined below, though, not necessary in our calculation. Note also that the two states closest to $k=\pi/a$ have strong edge state character. On the other hand, for the constriction the dispersion relation is shown in Fig.~\ref{Fig2}d. In the marginal VPM phase, each valley still has a direct gap but is shifted upward or downward due to the presence of $h$, resulting in a zero indirect gap. The two-terminal conductance of the QCP can be calculated by the Landauer formula
\begin{equation}
G=\frac{e^2}{h}\sum_{\mu\nu}|t_{\mu\nu}|^2\equiv\frac{e^2}{h}\left( T_{\uparrow}+T_{\downarrow} \right),
\end{equation}
where the spin-resolved transmission probability $T_{\uparrow(\downarrow)}=\sum_{m\in \uparrow(\downarrow)}\sum_{n}|t_{mn}|^2$ with $m$ and $n$ representing outgoing and incoming channels, respectively. The transmission matrix $t_{mn}$ can be computed numerically by the iterative Green's function method \cite{ando91}. The spin polarization can now be expressed as
\begin{equation}
P=\frac{T_{\uparrow}-T_{\downarrow}}{T_{\uparrow}+T_{\downarrow}}.
\end{equation}
For $0<P\leq 1$, the transmitted current is polarized with spin-up electrons, while for $-1\leq P<0$, the polarization is reversed.

For a specific demonstration, we consider a geometry with the length of the constriction $L_x=86a$, the width of the wide region $L_y=70\sqrt{3}a$, $L=34a$, and the width of the constriction $W=40\sqrt{3}a$; we adopt realistic parameters given in Table I for silicene and take $\lambda_{\text{v}}=E_{\text{c}}l$ for the wide regions to be in the SVPM phase, and $\lambda_{\text{v}}=0.053t$ $(>\lambda_{\text{SO}})$ and $h=0.05t$ for the gated constriction to be in the marginal VPM phase. We set $E_{\text{F}}=0.07t$, resulting in $2N=10$ right-moving modes. The resulting spin polarization as a function of the effective chemical potential, $\mu_0\equiv E_{\text{F}}-U_0$, is shown in Fig.~\ref{Fig2}b. Clearly, for positive $\mu_0<2h\approx 0.1$eV, the current flows entirely within the conduction band of the left valley and the polarization reaches almost 100\% efficiency with down-spin. Conversely, for negative $\mu_0>-2h$, the current flows within the valence band of the right valley and nearly 100\% spin-polarization is now achieved with up-spin. The case discussed here is the most optimal one. It is possible to use different $E_z$ and $h$ values in the constriction, which may turn silicene into other phases, but we find that the spin filter still remains functional (See Supplementary Figs.~S3-S5).


{\bf Transport properties.} In order to extract which transport features are the key ingredients for the high-efficiency spin polarization in our proposed device, it is instructive to consider a simple transport arrangement which bisects a long strip of silicene (with zigzag edges) into two halves as shown in Fig.~\ref{Fig3}a. We now apply out-of-plane electric fields $E_1$ and $E_2$ to the left and right regions, respectively. For simplicity, we assume the Zeeman field $h=0$ hereafter. The robust helical edge states (which cross from the conduction band to the valence band in the QSH phase as $E_i<E_{\text{c}}$) may or may not be present depending on the strength of the fields. Let us examine the case, $0<E_{\text{c}}<E_1$ with $E_2=-E_1$ (See Fig.~\ref{Fig3}b). Making $E_2$ negative but still equal to $E_1$ does not change the energy dispersion, but switches the spin configurations around the two valleys. Our calculation of the conductance and spin polarization (See Supplementary Table S1) indicates that for one specific spin polarization (e.g., up-spin), intervalley scattering is still partially allowed, although with serious suppression of the total conductance. But, spin-flip scattering becomes entirely forbidden. This is in sharp contrast to the classic example of the valley filter in graphene where valley index is preserved during the scattering process due to special features of the well-known single-valley edge mode \cite{valley2,fujita96,nakada96}. Remarkably, in our proposed spin filter the {\it bulk} transport from the coupled valley and spin degrees of freedom is the key instead of the edge states. In other words, bulk states with non-spin flip scattering processes are the crucial ingredients for the high efficiency spin polarization in our case.


\section{Discussion}

Our high efficiency, field tunable spin filter based on silicene takes advantage of charge carriers in the bulk system, small Rashba SOC, and controllable spin splitting due to IS breaking, $\lambda_{\text{v}}$. We expect therefore our spin-filter to be robust against weak disorder (compared to $\lambda_{\text{SO}}+h$) and edge imperfections. To test this expectation, we have added random onsite potential with strength less than $0.1t$ ($\sim h$) to the constriction. The polarization diagram is found to be almost quantitatively the same as that shown in Fig.~\ref{Fig2}b. In particular, suppression of the polarization is small, less than 2\% as $|\mu_0|<2h$. Furthermore, we have also changed the connection area of the constriction to the wide regions to be armchair-like, and introduced random vacancies around the edge of the constriction (up to 20\% of edge atoms removed), and found that the degradation of polarization is again not severe (less than 1\%) for such perturbations (See Supplementary Figs.~S5-S7 and Supplementary Discussions). 

The working temperature of our proposed silicene spin filter will be controlled by the parameter 
$2\lambda_{\text{SO}}=8.4$ meV, yielding a temperature of 97K which lies above the boiling point of liquid nitrogen. A similar germanene based device with $2\lambda_{\text{SO}}=23.6$ meV could be operated at room temperature. 
As to the characteristic gate voltage, we used a fairly large value of 1.57 V/$\text{\AA}$ for $E_z$ in the present simulations corresponding to $h\approx 0.052$ eV in the narrow constriction. 
Such a sizable exchange field could nevertheless be induced by the magnetic proximity effect as shown in the case of graphene.\cite{ding11} Using a few times smaller value of $E_z$ with a corresponding smaller $h$ (while maintaining the marginal VPM phase) will still keep our spin-filter functional, although the efficiency and the working region for $\mu_0$ could be moderately suppressed due to decrease of spin-splitting in the subbands. We emphasize however that such reduction in efficiency could be minimized by elongating the constriction, as indicated by the simulations of Supplementary Table S2.

Our spin filter device derives its unusual transport properties via the 2D features of silicene under IS breaking and interplay between the electric and magnetic fields. To further elucidate the uniqueness of silicene and its potential for future spintronics applications, we discuss the Y-shaped spin/valley separator shown in Fig.~\ref{Fig4}. The purpose of the device is to separate the two spin/valley polarizations from the incoming lead 1, with one running to lead 2 and the other running to lead 3. This separator could be operated as follows: First turn on an out-of-plane electric field, $E_z>E_{\text{c}}$, in the central silicene sheet, tune chemical potential $\mu$ into conduction bands, and then apply an in-plane electric field by setting potentials, e.g., $V_1>V_2=V_3$, at the terminals of silicene. This setup causes charge carriers to acquire an anomalous velocity proportional to the Berry curvature in the transverse direction, similar to that found out by Xiao {\it et al.} in graphene \cite{xiao07}. By linear response theory with negligible $\lambda_{\text{R}}$, the Hall conductivity around valley K$_{\eta}$ ($\eta=\pm 1$ for two valleys) with spin polarization $\sigma$ is given by
\begin{equation}
\sigma^{\eta,\sigma}_{\text{H}}=\eta\frac{e^2}{2h}(1-\frac{\lambda_{\text{v}}-\eta\sigma\lambda_{\text{SO}}}{\mu}).
\end{equation}
In other words, such a setup leads to valley and hence spin polarization imbalance at output terminals $V_2$ and $V_3$ (with opposite polarization between them), and results in non-vanishing valley Hall [$\sigma_{\text{H}}^{\text{(valley)}}=\frac{2e^2}{h}(1-\frac{\lambda_{\text{v}}}{\mu})$] and spin Hall conductivity [$\sigma_{\text{H}}^{\text{(spin)}}=\frac{2e^2}{h}(\frac{\lambda_{\text{SO}}}{\mu})$], respectively. Detailed demonstration of this separator requires delicate simulations, which will be considered elsewhere. Our study demonstrates that silicene (and related IVA group elements with honeycomb structure \cite{liu11a}) provides a great potential host for manipulating spin/valley degrees of freedom efficiently, moving us a step closer to realizing the dream of spintronic/valleytronic applications.

\section{Methods}

Our first-principles calculations are based on the generalized gradient approximation (GGA) \cite{PBE}
using full-potential projected augmented wave method \cite{PAW} as
implemented in the VASP package \cite{VASP}.
The 2D low-buckled honeycomb structures of Si, Ge, Sn, and Pb were optimized using a 30 $\times$ 30 $\times$ 1
Monkhorst-Pack k-point mesh over the Brillouin zone (BZ) with 350 eV cutoff energy.
 We note that our computed value of band gap of 8.4 meV in silicene is close to the value of 7.9 meV reported in Ref.~\onlinecite{liu11b}. 
The transport simulations are based on the iterative Green's function method \cite{ando91} (See Supplementary Methods).

{\bf Refereneces}

{\bf Acknowledgements}

WFT is grateful to the hospitality of Kavli Institute for Theoretical Physics China, CAS, China, where part of the work was done. WFT and CYH are supported by the NSC in Taiwan under Grant No. 100-2112-M-110-001-MY2. TRC and HTJ are supported by the National Science Council and Academia Sinica, Taiwan. 
The work at Northeastern University is supported by the Division of Materials Science and Engineering, Basic Energy Sciences, US Department of Energy through grant number DEFG02-07ER46352, and benefited from
the allocation of supercomputer time at NERSC and Northeastern University Advanced Scientific Computation Center and theory support at the Advanced Light Source (through grant number DE-AC02-05CH11231). 
We also thank NCHC, CINC-NTU, and NCTS, Taiwan for technical support.

{\bf Author Contributions}

All authors contributed extensively to the work presented in this paper.

{\bf Competing Financial Interests statement}

The authors declare no competing financial interests.

*Correspondence and requests for materials should be addressed to
H.L. (Email: nilnish@gmail.com).

\newpage

\begin{table}
\caption{{\bf Model parameters.} Fitting results from first-principles calculations for stand-alone, 2D Si, Ge, Sn, and Pb sheets.
Relevant parameters are lattice constant ($a$) with buckled structure, buckling distance ($\Delta_{\text{c}}$), transfer energy ($t$), spin-orbit coupling ($\lambda_{\text{SO}}$), Rashba spin-orbit coupling ($\lambda_{\text{R}}$), and the linear dependence of the applied electric field on $\lambda_{\text{v}}$ ($l\equiv\lambda_{\text{v}}/E_z$).}
\label{table1}
	\begin{center}
		\begin{tabular}{cccccccc}
\hline \hline
& $a$    & $\Delta_{\text{c}}$ & $t$     & $\lambda_{\text{SO}}$ & $\lambda_{\text{R}}$ & $l$   \\
& (\AA)& (\AA)        & (eV) & (meV)          & (meV)         & (e$\cdot$\AA)  \\
\hline
Si	&	3.87	&	0.44	&	1.04	&	4.2	&	8.66	&	0.035	\\
Ge	&	4.06	&	0.68	&	0.97	&	11.8	&	2.81	&	0.046	\\
Sn	&	4.67	&	0.84	&	0.76	&	36.0	&	18.75	&	0.055	\\
Pb	&	4.93	&	0.89	&	0.72	&	207.3	&	0.06	&	0.143	\\
\hline \hline
		\end{tabular}
	\end{center}
\end{table}

\newpage

\begin{figure}
\caption{{\bf Nearly fully spin-polarized states of silicene.}
{\bf a}, Low-buckled 2D honeycomb structure of silicene. Due to the buckling, the inversion symmetry can be removed by an external out-of-plane electric field $E_z$ or when the thin film is placed on a substrate.
{\bf b}, Schematic spin-resolved band structure of silicene around K and K' points in the presence of an out-of-plane electric field, $E_z<E_{\text{c}}$. The red and blue arrows indicate the spin direction.
The band structure
({\bf c}) and spin-resolved wavefunction
({\bf d}) in the critical phase ($E_z=E_{\text{c}}$).
The tight-binding model (green lines) is seen to faithfully reproduce the band dispersion and the nearly 100\% electron intrinsic spin-polarization of states near the K-point obtained by first-principles calculations (red dots).
{\bf e}, The spin splitting energy for silicene, and Ge, Sn, Pb counterparts as a function of $E_z$ obtained by first-principles calculations.
}
\label{Fig1}
\end{figure}

\newpage

\begin{figure}
\caption{{\bf Tunable high efficiency spin filter.}
{\bf a}, Geometry of the spin filter and the profile of the potential barrier $U(x_i)$. The two colored atoms on the lattice emphasize the buckled structure.
{\bf b}, Spin polarization of the filter as a function of $\mu_0$ in the constriction. The blue (green) line corresponds to the case of applying potential barrier with a rectangular (smooth) shape (see text).
{\bf c} and {\bf d} are typical dispersion relations for the wide and the constriction regions, respectively. 
}\label{Fig2}
\end{figure}

\newpage

\begin{figure}
\caption{{\bf Transport selectivity due to spin and valley degrees of freedom.}
{\bf a}, Transport between two separated regions (by applying different electric fields) of a silicene strip. {\bf b}, A schematic illustration for robust spin-preserved (instead of valley-preserved) scattering processes.
}
\label{Fig3}
\end{figure}

\newpage

\begin{figure}
\caption{{\bf Y-shaped silicene spin separator.} A schematic Y-shape separator of silicene with three local gates (leads). The current separates into the gate 2 and gate 3, respectively, carrying opposite spin/valley degrees of freedom.}
\label{Fig4}
\end{figure}

\end{document}